\documentclass[final,5p,times,twocolumn]{elsarticle}
\usepackage{graphicx}
\usepackage{amsmath}
\usepackage{amssymb}
\usepackage{bm}
\usepackage{mathrsfs}
\usepackage{color}
\usepackage{slashed}
\usepackage{dcolumn}

\begin{document}

\begin{frontmatter}

%\title{Inclusive jet and di-jet production at $ {\cal O}( \alpha_s^3 )$ in heavy
%ion reactions at the LHC}

\title{ ${\cal O}(\alpha_s^3)$ analysis of inclusive jet and di-jet production in heavy
ion reactions \\
at the Large Hadron Collider }

 \author[]{Yuncun He$^{\;a}$}
 \ead{heyc@iopp.ccnu.edu.cn}
\address[]{Key Laboratory of Quark and Lepton Physics (Central China Normal University),
Ministry of Education, China}
\address[]{Los Alamos National Laboratory, Theoretical Division, MS B238, Los Alamos, NM 87545, U.S.A.}
 \author[]{Ivan Vitev$^{\; b}$}
\ead{ivitev@lanl.gov}
\author[]{Ben-Wei Zhang$^{\;a}$}
\ead{bwzhang@iopp.ccnu.edu.cn}

\begin{abstract}
Jets physics in heavy ion  reactions is an important new area of active research at the
Relativistic Heavy Ion Collider (RHIC) and at the Large Hadron Collider (LHC) that paves the way
for novel tests of QCD multi-parton dynamics in dense nuclear matter. At present, perturbative
QCD calculations of hard probes in elementary nucleon-nucleon reactions can be consistently
combined
with the effects of the nuclear medium up to $ {\cal O}( \alpha_s^3 ) $. While such accuracy
is desirable but not necessary
for leading particle tomography, it is absolutely essential for the new jet observables.
With this motivation, we present first results and predictions to
$ {\cal O}( \alpha_s^3 ) $ for the recent LHC lead-lead (Pb+Pb) run at a
center-of-mass energy of 2.76~TeV per nucleon-nucleon pair. Specifically, we focus on
the suppression of
the single and double inclusive jet cross sections. Our analysis includes not only
final-state inelastic parton interactions in the QGP, but also initial-state cold nuclear
matter effects and an estimate of the non-perturbative hadronization corrections.
We demonstrate how an enhanced di-jet asymmetry in central Pb+Pb reactions at the LHC,
recently measured by the ATLAS and CMS experiments, can be derived from these results.
We show quantitatively that a fraction of this enhancement may be
related to the ambiguity in the separation between the jet and the soft background medium
and/or the diffusion of the parton shower energy away from the jet axis through
collisional processes. We point to a suite of measurements that can help build a
consistent picture of parton shower modification in heavy ion collisions at the LHC.
\end{abstract}

%\begin{keyword}

%\end{keyword}

\end{frontmatter}

% \linenumbers

\section{Introduction}
\label{introduction}

One of the most intensively studied processes in high-energy physics is jet production,
which can be used to accurately test the fundamental properties of Quantum Chromodynamics
(QCD) and explore new physics beyond the Standard Model~\cite{Campbell:2006wx}. At the parton level,
jet observables, such as cross sections, jet shapes or event shapes, can be computed
systematically and reliably order-by-order in perturbative Quantum Chromodynamics (pQCD).
At the hadron level, non-perturbative effects related to hadronization and the underlying event
can be estimated and combined with the pQCD results to produce  theoretical prediction
suitable for comparison to the experimental
measurements~\cite{Campbell:2006wx,Olness:2009qd,Dasgupta:2007wa,Cacciari:2008gd}.
The highest center-of-mass energy hadronic collisions available to date at the CERN  Large
Hadron Collider (LHC) guarantee an abundant yield of large transverse momentum  hadrons and jets
and provide, through experimental measurements and constraints on theoretical
models~\cite{:2010wv,Khachatryan:2011tk}, an ideal opportunity to advance perturbative QCD.

In relativistic heavy ion collisions at the LHC, jet and particle production will
be altered by the collisional and radiative final-state interactions of the parent partons
which propagate through a region of hot and dense deconfined matter, or the quark-gluon
plasma (QGP). It has been demonstrated that a significant part of the partons' energy
can be lost due to medium-induced gluon bremsstrahlung~\cite{jet1}, and uncertainties in
energy loss calculations have been commented upon in~\cite{Majumder}. The predicted related
suppression of leading particles at the Relativistic Heavy Ion Collider (RHIC)  and at
the LHC~\cite{Vitev:2005he} is now well
established~\cite{:2008cx,Abelev:2009wx,Aamodt:2010jd}, including its non-trivial momentum
dependence verified by the ALICE Collaboration. On the other hand, the modification of jets
in dense QCD matter requires much deeper understanding of the in-medium parton shower
production mechanisms~\cite{Vitev:2008rz}. For inclusive jets, tree-level calculations
can be used to discuss qualitatively the effects of final-state interactions but
$ {\cal O}( \alpha_s^3 ) $ results~\cite{jets} are necessary for quantitative comparison to
the preliminary jet experimental data in gold-gold (Au+Au) and copper-copper (Cu+Cu) reactions at
RHIC~\cite{Salur:2010qk,Jia:2010hg,Ploskon:2009zd}. For tagged jets and di-jets
lowest order, for example ${\cal O}(\alpha_s^2)$, ${\cal O}(G_F\alpha_s)$ and
  ${\cal O}(\alpha_{\rm em} \alpha_s)$,  calculations fail to describe even qualitatively the double
differential momentum distributions~\cite{Neufeld:2010fj} and one-loop pQCD results
are required.

Recently the ATLAS and CMS Collaborations reported  a significant enhancement
in the transverse momentum imbalance of di-jets produced  in central Pb+Pb
collisions at the LHC~\cite{Aad:2010bu,Chatrchyan:2011sx}. It has been qualitatively
argued that such asymmetry $A_J$  may be a consequence of jet quenching in the QGP
created in heavy ion reactions at LHC~\cite{CasalderreySolana:2010eh}. Attempts
to explain its magnitude based on Monte Carlo simulations with a Pythia generated
p+p baseline have also been presented~\cite{Qin:2010mn,Lokhtin:2011qq,Young:2011qx}.
In  Ref.~\cite{Cacciari:2011tm} the important observation was made that the huge heavy ion
background and its fluctuations may significantly enhance the di-jet asymmetry
distribution.

In this Letter we report first results and predictions at $ {\cal O}( \alpha_s^3 ) $ for
inclusive jet and di-jet productions in central Pb+Pb collisions with $\sqrt{s_{NN}}=2.76$~TeV
at the LHC. We combine consistently the $ {\cal O}(\alpha_s^3)$ calculations
in nucleon-nucleon reactions, validated through comparison to the $\sqrt{s}=7$~TeV
p+p data at the LHC,  with  parton energy loss effect in the hot QCD medium
at $ {\cal O}(\alpha_s^2 \alpha_s^{\rm rad}) $~\cite{Vitev:2010ci}.
$ {\cal O}(\alpha_s^3)$, $ {\cal O}(\alpha_s^2 \alpha_s^{\rm rad}) $
results represent next-to-leading order accuracy for inclusive jet production
but for the more differential observables, such as the di-jet asymmetry at
$A_J>0$,  this may be the first order that gives a non-trivial ($\neq 0$) contribution.
We demonstrate the relation between the cross section attenuation for inclusive jets and
di-jets and the resulting enhancement in the di-jet asymmetry $A_J$.
We find that a  part of this enhancement may arise from the uncertainty in the
jet/background separation and/or the dissipation of the energy from the shower
into the QGP due to collisional processes~\cite{source}. To further differentiate these
two possibilities is subject of future work.

Our Letter is organized as follows: we present benchmark results for inclusive jet and di-jet
production in $\sqrt{s}=7$~TeV  p+p collisions at the LHC in section~\ref{ppformalism}.
Results and predictions for the central Pb+Pb reactions at the LHC at $\sqrt{s_{NN}} = 2.76$~TeV
are given in section~\ref{AAformalism}. Our conclusions are presented in section~\ref{conclude}.

\section{Inclusive jet and di-jet production in p+p collisions}
\label{ppformalism}
%%%%%%%%%%%%%%%%%%%%%%%%%%%%%%%%%%%%%%%%%%

Within the framework of the collinear perturbative QCD factorization
approach the lowest order (LO) single and double inclusive jet production cross
sections can be obtained from:
\begin{eqnarray}
\!\!\!\!\!\!\!\!\!    \frac{ d\sigma }{ dy_1  dy_2 d^2E_{T\,1}  d^2E_{T\, 2} }
&=&
 \sum_{ab,ij}  \frac{\delta (\Delta \varphi_{1,2} - \pi)
\delta (E_{T\,1}- E_{T\,2} ) }{ E_{T\,2} } \nonumber \\
&& \hspace*{-2.5cm}
\times  \, \frac{\phi_{a/N}({x}_a) \phi_{b/N}({x}_b) }{{x}_a{x}_b}  \,
\frac{\alpha_s^2}{{s}^2 } \, |\overline {M}_{ab\rightarrow ij}|^2  \;.
\label{basic}
\end{eqnarray}
In  Eq.~(\ref{basic})  $ s = (P_a + P_b)^2 $  is the  squared center-of-mass energy of
the hadronic collision and $ d^2E_T=  E_TdE_Td\varphi$ with $\varphi$ being the
azimuthal angle of the jet.
$$x_a = \frac{p_a^+}{P_a^+} =\frac{E_{T\,1}}{\sqrt{s}}(e^{y_1}+e^{y_2})
\, , \;   x_b = \frac{p_b^-}{P_b^-} = \frac{E_{T\,2}}{\sqrt{s}}(e^{-y_1}+e^{-y_2}) \,  $$
are  the lightcone momentum fractions of the incoming partons. We denote
by $ \phi_{a,b/N}(x_{a,b})$  the  distribution function of
partons $a,b$ in the nucleon. $ |\overline {M}_{ab \rightarrow ij}|^2 $
are the squared  matrix elements for  $a+b \rightarrow i+j $ processes, for massless partons
see~\cite{Owens:1986mp}. In the double collinear approximation the jets are produced
back-to-back and identified with the parent partons. Formally, the partonic cross section in
Eq.~(\ref{basic}) is convolved with the jet phase space $S_2(p_1^{\mu},p_2^{\mu})$
(dimensionless in our convention). At tree level,
this phase space factor is simply a sum of products of $\delta$-functions.
For  inclusive jet production,  the second  jet is integrated out.

%%%%%*************************------------------------------------
\begin{figure}[!t]
\vskip0.04\linewidth
\centerline{
\includegraphics[width = 0.9\linewidth]{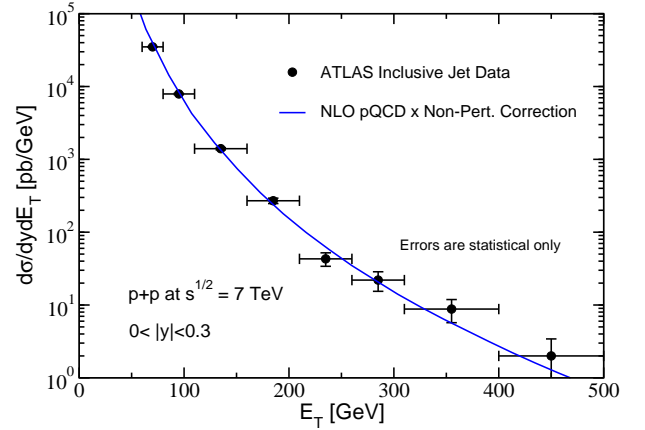}
}
\caption{ Comparison between the $R=0.4$ inclusive jet cross section measured by the
ATLAS Collaboration
in p+p collisions at $\sqrt{s}=7$~TeV  and the $ {\cal O}( \alpha_s^3 ) $ QCD theory which includes non-perturbative
corrections. Experimental error bars are statistical only.}
\label{tevdata}
\end{figure}
%%%%%%%%*********************************************

The first obvious deficiency of the LO formalism is its inability to describe the jet radius
dependence of the physical observables~\cite{jets}. The second much more significant shortcoming is
the inadequate description of the differential distribution of tagged jets~\cite{Neufeld:2010fj} and
di-jets due to the $\delta$-function constraint in  Eq.~(\ref{basic}).
At $ {\cal O}( \alpha_s^3 ) $ and beyond jets are defined by the jet-finding algorithms characterized by
a  radius parameter $R=\sqrt{\Delta y^2+\Delta \phi^2}$, which gives the jet's spacial extent in
rapidity $y$ and azimuthal angle $\phi$. Generally, the cross sections of the jet production at the $ {\cal O}( \alpha_s^3 ) $
can be expressed as~\cite{EKS}:
\begin{eqnarray}
 \!\!\!\! \!\!\! \frac{d\sigma}{dV_{\rm n}}&=& \frac{1}{2!}\int dV_{\rm 2}
\frac{d\sigma(2\rightarrow2)}{dy_1dy_2d^2E_{T\, 1}d^2E_{T\,2} }S_2(p_1^{\mu},p_2^{\mu})\nonumber \\
 &&  +\frac{1}{3!}\int dV_{\rm 3} \frac{d\sigma(2\rightarrow3)}
{dy_1dy_2dy_3d^2E_{T\, 1}d^2E_{T\,2} d^2E_{T\,3} }
\nonumber \\
&& \hspace*{1.3cm}\times \,S_3(p_1^{\mu},p_2^{\mu},p_3^{\mu}) \,\, ,
   \label{di-pt}
\end{eqnarray}
where $V_{\rm n} = dy_1\cdots dy_nd^2E_{T\,1}\cdots d^2E_{T\,n}$ represents the  phase space for
 multi-jets or multi-partons, respectively.
The first term on the right-hand side of Eq.~(\ref{di-pt}) gives the contribution
from $2 \rightarrow 2$ processes, including the $ {\cal O}( \alpha_s^3 ) $ virtual corrections.
The second term in Eq.~(\ref{di-pt}) denotes the contribution from $2 \rightarrow 3$ processes.
The jet size $R$ and algorithm dependence are contained in the function
$S_3(p_1^\mu,p_2^\mu,p_3^\mu) $~\cite{jets}, which is significantly more complicated than
$S_2(p_1^\mu,p_2^\mu) $. In the calculations that follow we will take
advantage of the Ellis, Kunszt and Soper (EKS) numerical implementation of
$ {\cal O}( \alpha_s^3 ) $ jets production in hadron-hadron collisions~\cite{EKS} and the CTEQ6.1M parametrization
set of parton distribution functions~\cite{Pumplin:2002vw}.

Measurements of inclusive jet production in p+p collisions at $\sqrt{s} = 7$~TeV with
jet cone radii $R=0.4, 0.6$  and in different rapidity regions have been carried out by the
ATLAS Collaboration at the LHC~\cite{:2010wv}. Perturbative QCD calculations can be
corrected to the hadron level using parametrization of universal non-perturbative effects $f_{\rm NP}(E_{T},R)$
extracted from Monte Carlo simulations as follows:
\begin{equation}
\frac{d\sigma^{\rm hadron}}{dE_{T\,1}\cdots dE_{T\,n}}
= \frac{d\sigma^{\rm parton}}{dE_{T\,1}\cdots dE_{T\,n}}\prod_{i=1}^n f_{\rm NP}(E_{T\,i},R_i) \, .
\label{NP}
\end{equation}
In this Letter we employ the parametrization of non-perturbative effects provided
by ATLAS~\cite{:2010wv} in all calculations except the ones presented in the bottom
panel of Fig.~\ref{PTnoPT}. The only observables that are noticeably affected
by  $f_{NP}$ are the ones that involve too significantly different radii $R_1, \, R_2$.
Returning to jet production in p+p collisions, we find that
an excellent description of the differential jet  spectrum
is obtained in the measured region of $E_T$ between 50~GeV and 500~GeV where the
yield falls by more than 5 orders of magnitude. We present one example for
rapidity $|y|<0.3$ and $R=0.4$ in Fig.~\ref{tevdata}. Variation of the factorization and
renormalization scale $\mu_f=\mu_R = E_T$ in the interval $(E_T/2, 2E_T)$ has less than
10\% effect on the inclusive jet cross section.

\begin{figure}[!t]
\vskip0.04\linewidth
\centerline{
\includegraphics[width = 0.85\linewidth]{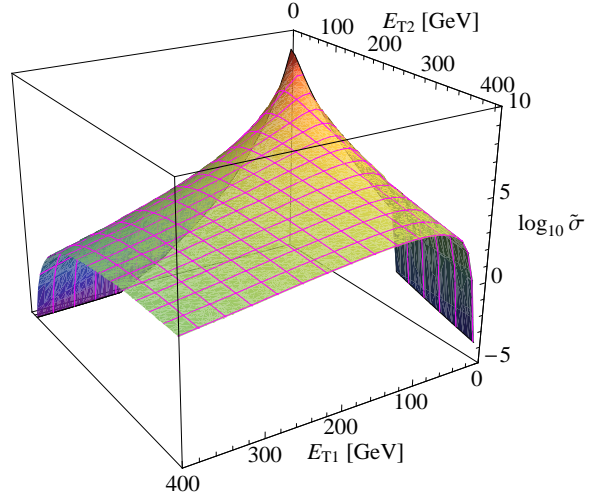} }
\caption{A three-dimensional representation of the $ {\cal O}( \alpha_s^3 ) $  di-jet cross section
$\tilde{\sigma}$, defined in Eq.~(\ref{brief}), in $\sqrt{s} = 7$~TeV p+p
collisions at the LHC  versus $E_{T\, 1}$ and $E_{T\, 2}$. We have used a jet
radius $R=0.4$ and a rapidity interval $|y|<2.8$. }
\label{3D-e1e2}
\end{figure}

The evaluation of di-jet production in hadronic reactions is slightly more involved.
At $ {\cal O}( \alpha_s^3 ) $ the naive exact transverse energy equality $E_{T\, 1} =  E_{T\, 2}$
between the two jets is broken by  $2 \rightarrow 3$ parton splitting  processes.
Loop corrections in $2 \rightarrow 2$ processes, proper
jets definitions and finite $E_T$ bins ensure well-behaved physical cross sections.
To quantify the resulting imbalance we modify the $ {\cal O}( \alpha_s^3 ) $ EKS
code~\cite{EKS} and define the dimensionless:
\begin{equation}
\tilde{\sigma} =  \left[\frac{1}{nb}\right] \,
E_{T\,1}E_{T\,2} \frac{d\sigma}{dE_{T\,1}dE_{T\,2}}   \;
\label{brief}
\end{equation}
for brevity. In Fig.~\ref{3D-e1e2} we show the three-dimensional (3D)  plot of
$ \tilde{\sigma} $ as a function of the transverse energy of the two jets $E_{T\, 1}$
and $E_{T\, 2}$ in  the rapidity region $|y|<2.8$ for the same jet size  $R=0.4$.
Transverse energy bins $\Delta E_{T\,1}, \Delta E_{T\,2} = 10$~GeV have been chosen.
While the cross section reaches its maximum for
$E_{T\, 1} \approx E_{T\, 2}$, its most striking feature is how broad and slowly varying the
di-jet yield is away from the main diagonal. As we will see below, such broad distribution
limits the additional asymmetry that jet propagation in the QGP will induce.

%%%%%%%%%-------------------------------
%%%%%%%%%-------------------------------

One particular distribution that can be extracted from our general di-jet cross section
calculation is that of the di-jet asymmetry $A_J$, defined   as
\begin{equation}
A_J=\frac{E_{T\,1}-E_{T\, 2}}{E_{T\,1}+E_{T\, 2}} \;.
\label{ajdef}
\end{equation}
Making a change of variables from $(E_{T\,1}, E_{T\,2})$ to $(A_J,E_{T\,2})$ with $E_{T\,1} = E_{T\,2}(1+A_J)/(1-A_J)$
and then integrating over $y_1, y_2, E_{T\,2}$ we can express the $A_J$ distribution from the
di-jet $E_{T\,1},\, E_{T\,2}$
spectrum as follows:
\begin{eqnarray}
\frac{d\sigma}{dA_J}&=& \int_{y_{1\,\min}}^{y_{1\,\max}} dy_1 \int_{y_{2\,\min}}^{y_{2\,\max}} dy_2
\int_{E_{T\,2\, \min}}^{E_{T\,2\,\max}} dE_{T\, 2}  \; \nonumber \\[2ex]
&& \frac{2 E_{T\,2}}{(1-A_J)^2}\frac{d\sigma[E_{T\,1}(A_J,E_{T\,2})]}{dy_1dy_2 dE_{T\,1}dE_{T\,2}} \;,
\label{ajcalc}
\end{eqnarray}
where $E_{T\,2\, \max}$ is given by the jet production kinematics.  By evaluating
$A_J> 0$ we ensure that $E_{T\,1}>E_{T\,2}$. Note that at LO we always have $A_J = 0$.

%%%%%%%%%-------------------------------
\begin{figure}[t!]
\vskip0.04\linewidth
\centerline{
\includegraphics[width = 0.9\linewidth]{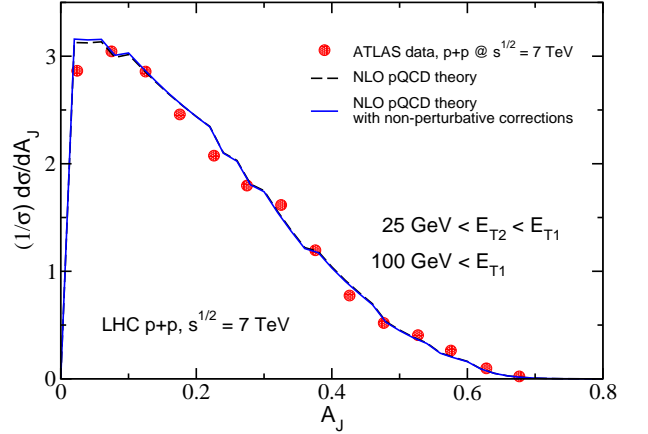}
}
\caption{
The normalized differential distribution of the di-jet asymmetry $A_J$, defined in
Eq.~(\ref{ajdef}), is evaluated at $ {\cal O}( \alpha_s^3 ) $ and compared to the ATLAS experimental measurement at
$\sqrt{s}=7$~TeV. Solid (dashed) lines include (do not include) non-perturbative
corrections, respectively. }
\label{ppAJ}
\end{figure}
%%%%%%%%%-------------------------------

With the $ {\cal O}( \alpha_s^3 ) $ numerical results for $\tilde{\sigma}$  at hand, we can easily compute
$\sigma^{-1} d\sigma/dA_J$ from Eq.~(\ref{ajcalc}) and confront it with the experimental data.
The ATLAS Collaboration has measured the di-jet asymmetry distribution
in p+p reactions at $\sqrt{s}=7$~TeV with the following kinematics cuts:
$E_{T\, 1}>100$~GeV for the leading jet  and $E_{T\, 2}>25$~GeV for the subleading
jet~\cite{Aad:2010bu}. The corresponding comparison to the perturbative QCD theory
is presented in Fig.~\ref{ppAJ} and shows excellent agreement.
 Variation of the factorization and
renormalization scale $\mu_f=\mu_R = E_T$ in the interval $(E_T/2, 2E_T)$ has less than
2\% effect on the $A_J$ distribution.
The dashed line represents the parton level $ {\cal O}( \alpha_s^3 ) $ results
and the solid line includes non-perturbative effects, which are seen to be practically
negligible for this observable. Strict transverse energy ordering  $E_{T\,2} < E_{T\,1}$
for the leading and subleading jets implies
$d\sigma/dA_J =0$ when $A_J=0$.

Fixed ${\cal O}(\alpha_s^3)$ calculations of $d\sigma/dA_J$  will diverge for $A_J \rightarrow 0$.
To ensure proper cancellation of real and virtual contributions to the physical cross section,
these evaluations rely on finite bin sizes in transverse energy, rapidity and azimuth.
For example, in numerical simulations based on the EKS code, results are calculated at random
positions in phase space and any required grid for the differential cross sections is
given by interpolation with a Gaussian smearing function~\cite{EKS}.
Thus, our numerical simulations yield continuously smooth results of  $E_{T\,1} \approx E_{T\,2}$
or, equivalently, $A_J \approx 0$.  On the other hand, fine resolution cannot be achieved when
$E_{T\, 1} \rightarrow E_{T\, 2} $ or $A_J \rightarrow 0 $. To investigate this region,
resummation beyond fixed order calculations will be
needed.~\footnote{We thank the reviewer for a suggestion to clarify this point.}

\section{Inclusive jet and di-jet production in heavy ion collisions}
\label{AAformalism}

The bases for the evaluation of multi-jet cross sections in heavy ion collisions are
the corresponding cross sections in the more elementary nucleon-nucleon reactions.
In Ref.~\cite{Ovanesyan:2011xy} it was shown that in QCD the final-state
process-dependent medium-induced radiative corrections factorize in the production cross sections
for the example of a single jet. The generalization to multiple jets proceeds as follows
\begin{eqnarray}
 d \sigma(\epsilon_1,\cdots,\epsilon_n)^{n-{\rm jet}}_{\rm quench.} &=&
d\sigma(\epsilon_1,\cdots,\epsilon_n)^{n-{\rm jet}}_{\rm pp}  \otimes P_1(\epsilon_1)
\cdots  \nonumber \\
\!\!\!\!\! && \otimes P_n(\epsilon_n)
\, \ |J_1(\epsilon_1)| \cdots |J_n(\epsilon_n)| \;. \qquad
\label{fact}
\end{eqnarray}
Here, $P_i(\epsilon_i)$ is the probability that the $i$th jet will lose a fraction
$\epsilon_i$ of its energy, $|J_i(\epsilon_i)|$ are phase space Jacobians, and
$\otimes$ denotes an integral convolution. Any dependence on the jet
reconstruction parameters is not shown explicitly in Eq.~(\ref{fact}). Furthermore,
possible small contributions of the medium-induced parton shower associated with one jet
to other jets that are very close in phase space $(y,\phi)$ are not included. For inclusive and
approximately back-to-back jets that we consider in this Letter, this effect is negligible.

The application of Eq.~(\ref{fact}) to inclusive jet production has been
discussed extensively in Refs.~\cite{Vitev:2008rz,jets}, where the angular distribution of the medium-induced
gluon radiation is computed using the Reaction operator approach~\cite{jet1,GLV}.
Here, we summarize the main
result. The medium-modified jet cross section per binary scattering is evaluated as
\begin{eqnarray}
&&\frac{1}{\langle  N_{\rm bin}  \rangle}
 \frac{d\sigma^{AA}(R)}{dy_1dE_{T\,1}} =  \sum_{q,g}
\int_{\epsilon_1=0}^1 d\epsilon_1 \;  P_{q,g}(\epsilon_1,E_{T\,1}) \nonumber \\
&& \times \frac{1}{ \left(1 - [1-f(R_1,p_{T\,1}^{\min})_{q,g}]  \epsilon_1\right)}
 \frac{d\sigma^{\rm CNM,NLO}_{q,g}(E_{T\,1}^\prime)} {dy_1 dE^{\, \prime}_{T\,1}} \; . \qquad
\label{incl}
\end{eqnarray}
In Eq.~(\ref{incl}) the phase space Jacobian reads
$$|J_i(\epsilon_i)| = 1/\left(1 - [1-f(R_i,p_{T\,i}^{\min})_{q,g}]  \epsilon_i\right)\, , $$
and  $E_{T\,i}^\prime = |J_i(\epsilon_i)|E_{T\,i}$ in the argument of $\sigma^{\rm CNM,NLO}_{q,g}$.
Note that in the expression above we have denoted by
\begin{eqnarray}
 f(R_i,p_{T\, i }^{\min})_{q,g}= \frac{\int_0^{R_i} dr
\int_{p_{T\,i}^{\min}}^{E_{T\,i}} d\omega \,
\frac{dI^{\rm rad}_{q,g}(i)}{d\omega  dr }}
{\int_0^{{R\, i}^{\infty}} dr \int_{0}^{E_{T\,i}} d\omega \,
\frac{dI^{\rm rad}_{q,g}(i)}{d \omega dr} } \; ,
\label{fraction}
\end{eqnarray}
the part of  the fractional energy loss $\epsilon_i$ that is redistributed inside the jet. Thus,
$f(R_i,p_{T\, i }^{\min})_{q,g}$ plays a critical role in defining the contribution
(or lack thereof  if $f(R_i,p_{T\, i }^{\min})_{q,g} \rightarrow 0 $)
of  the  medium-induced  bremsstrahlung  to  the  jet. Note that in Eq.~(\ref{fraction})
$p_{T\, i}^{\min}$ is a parameter that can be used to simulate processes that can alter the energy
of the jet beyond medium-induced bremsstrahlung. For example, by choosing
$p_{T\, i}^{\min} > 0$~GeV one  can investigate phenomenologically the uncertainty in
the jet/background separation and the diffusion of the parton shower energy away from
the jet axis due to collisional processes. Specifically, one can adjust the value of
$p_{T\, i}^{\min} $ to reflect the collisional energy loss of the parton shower
$\langle \delta E_T \rangle$ evaluated in Ref.~\cite{source}.

Cold nuclear matter (CNM) effects prior to the QGP formation,
such as initial-state energy loss, coherent power corrections and the Cronin
effect~\cite{Vitev:2008vk},
can be incorporated in the cross sections for quark and gluon jet production
$\frac{d\sigma^{\rm CNM, NLO}_{q,g}} {dy dE_T} $ and  their multi-jet generalization.
At the LHC we are interested in jets of $E_T > 20$~GeV and at these energy
scales of the effects listed above only initial-state energy loss, which is also
derived in the framework of the Reaction operator approach,  may play a
role~\cite{Vitev:2007ve,Neufeld:2010dz}. In this work we evaluate the effects of cold
nuclear matter on jet production in Pb+Pb collisions at the LHC  and comment on their
relevance to specific observables.

For di-jets produced in heavy ion reactions, the corresponding cross section
can be expressed as
\begin{eqnarray}
\!\!\!\! &&\!\!\!\! \frac{1}{\langle  N_{\rm bin}  \rangle}
\ \frac{d\sigma^{AA}(R)}{dy_1dy_2dE_{T\,1}dE_{T\,2} } =  \sum_{qq,qg,gg}
\int_{\epsilon_1=0}^1 d\epsilon_1 \int_{\epsilon_2=0}^1 d\epsilon_2
\nonumber \\
&& \!\!\!\! \frac{P_{q,g}(\epsilon_1,E_{T\,1})}{ \left(1 - [1-f(R_1,p_{T\,1}^{\min})_{q,g}]
 \epsilon_1\right)}
 \frac{P_{q,g}(\epsilon_2,E_{T\,2})}{ \left(1 - [1-f(R_2,p_{T\,2}^{\min})_{q,g}]  \epsilon_2\right)}
\nonumber \\
&& \!\!\!\! \times
 \frac{d\sigma^{\rm CNM,NLO}_{qq,qg,gg}(E_{T\,1}^{\, \prime}, E_{T\,2}^{\,\prime})}
{dy_1 dy_2 dE^{\, \prime}_{T\,1} dE^{\,\prime}_{T\,2} } \; . \qquad
\label{eq:JCS-AA}
\end{eqnarray}

The evaluation of the inclusive and di-jet cross sections at the LHC proceeds as follows:
the hard jet production processes are distributed in the transverse plane according to the
binary collision density $\sigma_{in} ( d  T_{AA}(b)/d{\bf x}_\perp) $. The density of the QGP
is taken to be proportional to the participant density $dN_{\rm part.}/d{\bf x}_\perp$.
Through isospin symmetry, necessary to account for the neutral particles, and parton-hadron
duality~\cite{Vitev:2008rz}  the gluon rapidity  density in central
Pb+Pb collisions at the LHC at $\sqrt{s_{NN}} = 2.76$~TeV is estimated to be $dN^g/dy = 2650$
from the ALICE experimental measurement of $dN^{ch}/d\eta$~\cite{Aamodt:2010pb}.
We take into account the
longitudinal Bjorken expansion of the medium and, by assuming local thermal equilibrium, we
can evaluate the physical quantities, such as the Debye screening scale, the in-medium jet
scattering cross sections, and the mean free paths~\cite{Vitev:2008rz,jets}, that are
employed in  the calculation of
the radiative energy loss~\cite{Vitev:2007ve}.

The energy loss calculation, which uses the collision geometry  described above,
includes kinematic constraints, and provides the fully differential distribution of the
medium-induced bremsstrahlung~\cite{Vitev:2007ve}, is computationally demanding and
performed separately. Using the analytic leading power dependence (up to logarithmic
corrections) of the energy loss on the coupling between the jet and the
medium $g_{\rm med }$
and the path of the parton propagation in the QGP:
\begin{equation}
\frac{dN^g}{ d \omega d^2{\bf k}_\perp}  \sim g_{\rm med}^4 \int_{t_0}^\infty t
 \rho({\bf x}_{0\perp} + {\bf n}_\perp t, t)  \, dt  \; ,
\label{line}
\end{equation}
we can precisely account for/study the sensitivity to the details of the in-medium
jet dynamics by scaling the averaged bremsstrahlung spectra.
Note that $g_{\rm med}^2/4\pi$ above should not be confused with $\alpha_s^{\rm rad}$ in the
medium-induced bremsstrahlung vertex.
In Eq.~(\ref{line})  ${\bf x}_{0\perp}$ is the jet  or di-jet production
point in the transverse plane and  ${\bf n}_\perp$ is the direction of its propagation.
Finally, we evaluate in real time the probability distribution for the parent partons
to lose  fractions of their energy $\epsilon_i = \sum_n (\omega_{n})_i / E_{ i}$ due to multiple
gluon emissions in the Poisson  approximation~\cite{Vitev:2005he}:
\begin{equation}
  \int_0^1P_{q,g}(\epsilon_i) d\epsilon_i = 1\, , \quad \int_0^1\epsilon_i
P_{q,g}(\epsilon_i) d\epsilon_i = \left\langle \frac{\Delta E_{i}}{E_i} \right\rangle_{q,g} \, . \quad
\label{peps}
\end{equation}
This approximation is consistent with the inclusive gluon emission calculations
currently used in jet quenching phenomenology~\cite{jet1,GLV}.

\subsection{Quenching of inclusive jets}

\begin{figure}[!t]
\vskip0.04\linewidth
\centerline{
\includegraphics[width = 0.9\linewidth]{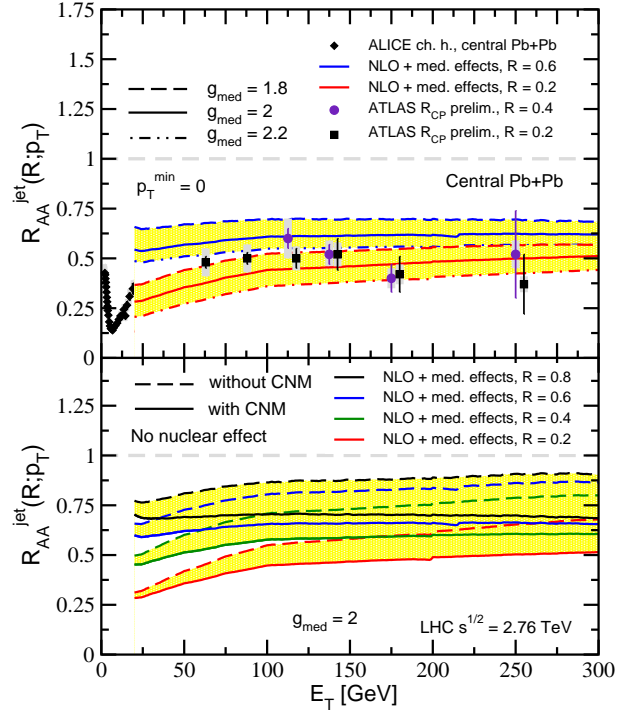}
}
\caption{ Top panel: the $E_T$ dependence of the nuclear modification factor for different
jet cone sizes  $R=0,2, \, 0.6$ is calculated in central Pb+Pb collisions at the
LHC $\sqrt{s_{NN}}=2.76$~TeV. Bands represent the variation in the coupling strength between
the jet and the medium. Bottom panel: the relative contribution of cold nuclear matter effects
to $R_{AA}$ is illustrated for $R=0.2, \, 0.4, \, 0.6, \, 0.8$. ALICE experimental data on
charged hadron suppression in central Pb+Pb collision is shown for reference. Preliminary ATLAS
$R_{CP}$ data is also shown for both $R= 0.2,\, 0.4$.}
 \label{RAAlhccoupling}
\end{figure}

To investigate jet production in relativistic heavy ion collisions and quantify
its deviation from the baseline results in elementary hadron-hadron reactions we
define a generalized nuclear modification factor for multi-jet final states
 as follows:
\begin{eqnarray}
&&R_{AA}^{\text{n-jet}}
(E_{T\,1}\cdots E_{T\, n} ; R_1 \cdots R_n , p_{T\, 1}^{\min}\cdots p_{T\, n}^{\min} ) \nonumber \\
&& =   \frac{
\frac{d\sigma^{AA}(E_{T\,1}\cdots E_{T\, n} ; R_1 \cdots R_n , p_{T\, 1}^{\min}\cdots p_{T\, n}^{\min} ) }
{ dy_1 \cdots dy_n    dE_{T\,1}\cdots dE_{T\, n} }
}
{ \langle  N_{\rm bin}  \rangle
\frac{d\sigma^{pp}(E_{T\,1}\cdots E_{T\, n} ; R_1 \cdots R_n , p_{T\, 1}^{\min}\cdots p_{T\, n}^{\min} ) }
{ dy_1 \cdots dy_n    dE_{T\,1}\cdots dE_{T\, n} }
  } \; .
\label{RAAjet}
\end{eqnarray}
Next, we calculate numerically the cross sections and nuclear modification
factors $R_{AA}^{\text{n-jet}}$  for inclusive jet and di-jet production ($n=1,2$)
in central  Pb+Pb collisions at LHC with $\sqrt{s_{NN}}=2.76$~TeV.

Fig.~\ref{RAAlhccoupling} illustrates the dependence of the suppression ratio $R_{AA}^{\rm 1-jet}$
for inclusive jets in central Pb+Pb collisions on the jet size $R$.
The top panel shows the net suppression resulting from both initial-state and final-state effects
for $R=0.2,\, 0.6$ and the bands represent the variation in coupling strength between the jet
and the medium.
The dashed, solid, and dot-dashed lines correspond to $g_{\rm  med}=$1.8, 2, 2.2, respectively.
Note that for small radii the suppression of jets can be comparable in magnitude to the
suppression of inclusive particles. This is illustrated by including a reference charged hadron
quenching measurement by the ALICE Collaboration~\cite{Aamodt:2010jd} in
Fig.~\ref{RAAlhccoupling}.  Preliminary ATLAS data~\cite{qmtalks} on $R_{CP}^{\rm 1-jet}$,
a ratio similar to $R_{AA}^{\rm 1-jet}$ where peripheral A+A reactions are used as a
baseline, has now been included in the
figure\footnote{Two weeks after this manuscript was been posted to the archive, the ATLAS
Collaboration showed preliminary results on inclusive jet quenching in central Pb+Pb collisions
at the Quark Matter 2011 conference. We include  two data sets  to give the reader a feel as to
how the predictions presented here compare to the new data.}.
The bottom panel of Fig.~\ref{RAAlhccoupling} presents a calculation for $R=0.2, \, 0.4, \, 0.6, \, 0.8$
and  for a fixed $g_{\rm  med}=$2  with (solid line) or without (dashed line) cold nuclear matter
effects. For fixed centrality, CNM effects, here dominated  by initial-state energy loss, do not
depend on the jet size or jet finding algorithm and become more relevant, relatively speaking,
for large radii $R$. Even though  on an absolute scale this  additional suppression is not very large, it
is more significant in comparison to the $Z^0$ or Dell-Yan production
processes~\cite{Neufeld:2010fj,Neufeld:2010dz,Chatrchyan:2011ua}.
These latter processes are dominated by $q+\bar{q}$ initial states and jet production
discussed in this manuscript  arises primarily from $g+g$ (and $g+q(\bar{q})$ at larger $E_T$) processes.

%%%%%%%%%%%%%****************************
\begin{figure}
\vskip0.04\linewidth
\centerline{
\includegraphics[width = 0.9\linewidth]{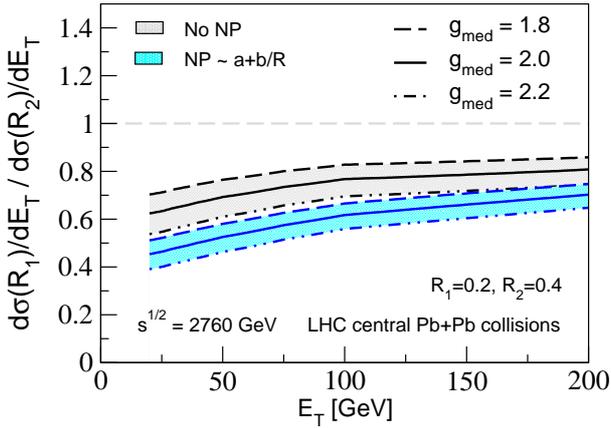}
}
\caption{ Ratio of the inclusive jet cross sections in central Pb+Pb collisions at LHC at
$\sqrt{s_{NN}}=2.76$~TeV for two different radii $R_1 =0.2$ and $R_2=0.4$. The bands show
results with different extrapolation of non-perturbative
corrections to small radii. The lines show effect of different coupling strength between the
jet and the medium. }
\label{R1toR2}
\end{figure}
%%%%%%%%%%%%%***************************

Initial-state CNM effects in heavy ion collisions can be minimized by taking the ratio of jet cross
section at two different radii  $[d\sigma(R_1)/dE_T]/[d\sigma(R_2)/dE_T]$~\cite{jets}. Since
the size $R$ determines what fraction of the  parton shower is reconstructed as a jet, it
affects the jet cross section. In heavy ion reactions  the cone size dependence is amplified by
the  fact that medium-induced parton showers have a broad angular distribution in comparison to the ones
in the vacuum~\cite{GLV}. This is shown  in Fig.~\ref{R1toR2} for $R_1=0.2, \, R_2=0.4$ and  the
dashed, solid, and dot-dashed lines correspond to three different $g_{\rm med} = 1.8, \, 2, \, 2.2$.
As the radius varies, specific
non-perturbative effects, unfortunately, become more important. Typically, they are expressed
as an average momentum shift~\cite{Olness:2009qd,Soyez:2011np} and related to ``splash-out'' hadronization
effects  and ``splash-in'' initial-state radiation/background
contribution: $\langle \delta p_T \rangle = A/R +  B R^2$. The physical effect of a momentum
shift is to alter the
measured cross section and this change can be isolated in a multiplicative factor~\cite{Vitev:2005he}.
Since background effects are the dominant uncertainty in jet physics with heavy ions,
we will discuss them separately.  With this motivation, we consider a
hadronization-motivated extrapolation  of the ATLAS parametrization
of non-perturbative effects to small radii: $f_{NP} = a + b/R$. The application of this
non-perturbative correction to the calculation of  $[d\sigma(R_1=0.2)/dE_T]/[d\sigma(R_2=0.4)/dE_T]$
in central Pb+Pb collisions at the LHC is shown by the cyan  band in  Fig.~\ref{R1toR2}, where we assume
that for jet production at large $E_T$ the "splash-out" hadronization effects in A+A collisions are the same as that in
elementary p+p collisions with the understanding that the hadronization time for jets at high $E_T$ is
longer than the duration time of the QGP.
Note that the non-perturbative effect can change significantly the cross section ratio
relative to the $ {\cal O}(\alpha_s^3) $ parton level result for small $R$. It is, therefore, critical to constrain
its magnitude as accurately  as possible in the simpler p+p reactions. Moreover, higher order corrections may
affect these ratios to some extent and a recent study
has shown that in p+p collisions higher order correction may lower the jet cross section ratios
considerably~\cite{Soyez:2011np}.

%%%%%%%%%%%%%****************************
\begin{figure}[!t]
\vskip0.04\linewidth
\centerline{
\includegraphics[width = 0.9\linewidth]{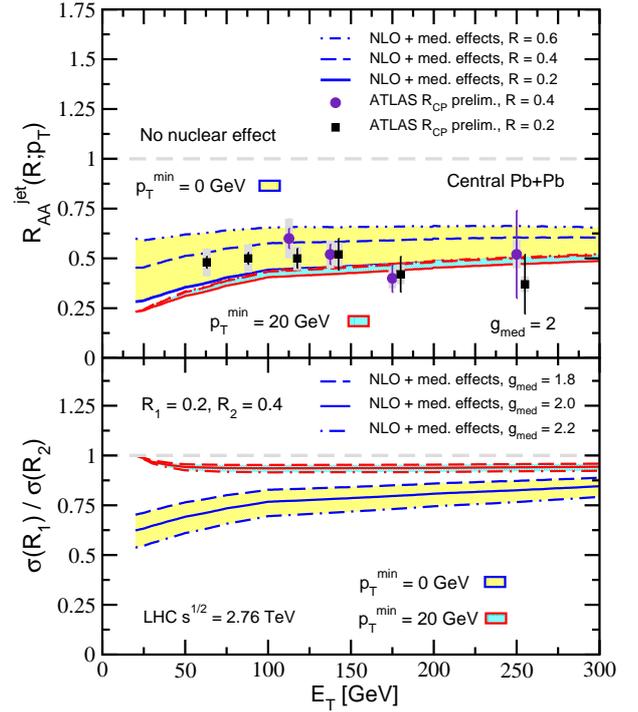}
}
\caption{ The effects of parton shower collisional energy loss or misinterpreting some
of the jet's energy as soft background on $R_{AA}^{\rm 1-jet}$
(top panel)   and $\sigma(R_1)/\sigma(R_2)$ (bottom panel) are simulated with $p_T^{\min} = 20 $~GeV and represented
by a cyan band. Results are contrasted to the $p_T^{\min} = 0$~GeV case, represented by a yellow band, which
is no collisional energy loss or  unambiguous jet/background separation.
Preliminary ATLAS $R_{CP}$ data is shown for $R=0.2,\, 0.4$. }
\label{PTnoPT}
\end{figure}
%%%%%%%%%%%%%***************************

Preliminary RHIC results suggest that the jet size dependence of jet attenuation may have already
been observed in Au+Au and Cu+Cu reactions at RHIC~\cite{Salur:2010qk,Jia:2010hg,Ploskon:2009zd}.
However, before we discuss di-jet production in heavy ion reactions, we  comment on the difficulties related
to the measurement of jet observables. In central Pb+Pb collisions at the LHC, for a typical jet size
$R=0.5$, on the order of 100~GeV of the jet energy is interpreted as QGP background and subtracted
from the total reconstructed energy~\cite{Chatrchyan:2011sx}.  While a simple jet+uniform background model
appears reasonable in heavy ion reactions, it is not based on first-principles theory. In what follows we
demonstrate the consequences of misinterpreting less than 20~GeV of the jet energy redistributed by
the QGP medium inside the jet as uncorrelated soft background. This is  $< 20\%$ of the
typical subtracted $E_T$ and in our approach~\cite{Vitev:2008rz} can be simulated by choosing
$p_T^{\min}=20$~GeV in Eq.~(\ref{fraction}).  We denote the actual amount of energy
by $\langle \delta E_T \rangle$ and emphasize that it depends on the jet radius $R$.
More importantly, we  recall that a recent calculation of the energy  transmitted by a
parton shower to the medium~\cite{source} $\Delta E({\rm shower \rightarrow QGP})$   found  that for LHC conditions
$\Delta E({\rm shower \rightarrow QGP}) \simeq 20$~GeV is well within reach, especially  for a gluon-initiated
shower. While the growth in the rate of collisional energy loss is due to the proliferation of
medium-induced partons, the rate for each individual parton is calculated to ${\cal O}(g_{\rm med}^4)$.
Of course, one should keep in mind that experimental observables are sensitive to the
physics effects on the fraction of the in-medium parton shower that falls within the cone radius $R$.
We first choose $p_T^{\min}=20$~GeV to simulate the collisional energy loss effects evaluated
in~\cite{source} on the parton shower for a very large radius $R \simeq 2 $. We then vary $R$
to obtain the collisional energy loss effect on the part of the parton shower that constitutes
the jet. For example, for a 100~GeV gluon jet choosing $p_T^{\min}=20$~GeV in
Eq.~(\ref{fraction}) leads to $ \langle \delta E_T \rangle \simeq 15$~GeV
for $R = 0.5$ and to  $\langle \delta E_T \rangle \simeq 5$~GeV for $R = 0.2$.
The effect of collisional energy loss decreases quickly with the jet radius since the medium-induced
parton shower has a characteristic  large-angle~\cite{GLV} distribution and most of it is already
outside of the jet cone  for small $R$~\cite{Vitev:2008rz,jets,Neufeld:2010fj}.

The result of our simulations is shown in Fig.~\ref{PTnoPT} for $g_{\rm med}=2$, where the default
choice  $p_T^{\min}=0$~GeV is illustrated by a yellow band and the choice $p_T^{\min}=20$~GeV - by a
cyan band. In the top panel, the clear dependence of $R_{AA}^{\rm 1-jet}$
on the jet size, exemplified by $R = 0.2,\, 0.4, \, 0.6$, practically
disappears if a sizable fraction of the particles created by the medium-induced
bremsstrahlung diffuse outside of the jet due to collisional processes or are  subtracted as
soft background. It can easily be seen that in this case and in the absence of strong $R$-dependent
non-perturbative corrections $\sigma(R_1)/\sigma(R_2) \approx 1$. This is demonstrated in the
bottom panel of Fig.~\ref{PTnoPT} for three $g_{\rm med} = 1.8,\, 2,\, 2.2$ and the result is very
different in comparison to letting $p_T^{\min}=0$~GeV. We do not include the non-perturbative
correction  in the bottom panel of Fig.~\ref{PTnoPT} since it can result in noticeably
different cross sections for significantly different $R_1,\, R_2$ and will distract from the
difference between the simulations that only include radiative energy loss processes and the ones
that include the effects of the collisional interactions of the parton shower.
We conclude that the ambiguity in the jet/background separation and the possibility that some of
the parton shower energy can diffuse away from the jet axis at angles greater than $R$ through
collisional processes~\cite{source} in heavy ion reactions can  significantly affect the experimental
jet cross sections.

\subsection{Quenching of di-jets and their $E_T$ asymmetry}
\begin{figure}
\vskip0.04\linewidth
\centerline{
\includegraphics[width = 0.95\linewidth]{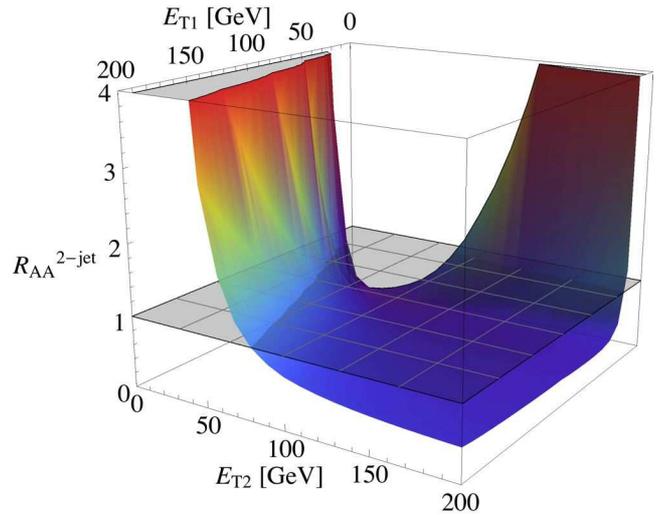}
}
\caption{$ {\cal O}( \alpha_s^3 ) $ calculation of the  di-jet cross section suppression ratio  $R_{AA}^{\rm 2-jet}$
versus $E_{T\, 1}$ and $E_{T\, 2}$  in  $\sqrt{s_{NN}}=2.76$~TeV central Pb+Pb collisions at the LHC. Our
result is for $R_1 = R_2 = 0.2$ and includes cold nuclear matter effects. }
\label{3DRdijet}
\end{figure}

We now present our results for the modification of di-jet production in central Pb+Pb collisions
at $\sqrt{s_{NN}} = 2.76$~TeV. The cross section is evaluated using Eq.~(\ref{eq:JCS-AA}). First, we give a
3D  representation for the suppression ratio $R_{AA}^{\rm 2-jet}$ of the di-jet cross section at $ {\cal O}( \alpha_s^3 ) $ in Pb+Pb
collisions relative to the theoretically computed  p+p  baseline in Fig.~\ref{3DRdijet}. In this example we
use jets with identical cone size $R=0.2$, a coupling strength between the jets and the medium
$g_{\rm med}=2$, and no transverse energy cut. The suppression involving both cold nuclear matter effects
and  final-state quenching effects in the QGP is the largest along the diagonal $E_{T\, 1}=E_{T\, 2}$.
In the regime $E_{T\, 1}\gg E_{T\, 2}$ or $E_{T\, 1}\ll E_{T\, 2}$ there is a striking enhancement
that can reach values much larger than 4 (the maximum value shown no the $z$ axis in Fig.~\ref{3DRdijet}).
Such enhancement was first identified theoretically on the example of $Z^0$ tagged jets~\cite{Neufeld:2010fj}
and is easy to understand. The final-state inelastic interactions of jets produced in heavy ion collisions
redistribute them to smaller transverse energies and these jets fill in a region of phase  space
where the cross section is steeply falling, see Fig.~\ref{3D-e1e2}. The second important observation
is how broad the region of approximately constant $R_{AA}^{\rm 2-jet}$ in the ($E_{T\,1},E_{T\,2}$)
plane is before it hits the  strong di-jet enhancement. This behavior limits the features in the
heavy ion $A_J$ enhancement that can be attributed to the physics of jet quenching.

\begin{figure}
\vskip0.04\linewidth
\centerline{
\includegraphics[width = 0.9\linewidth]{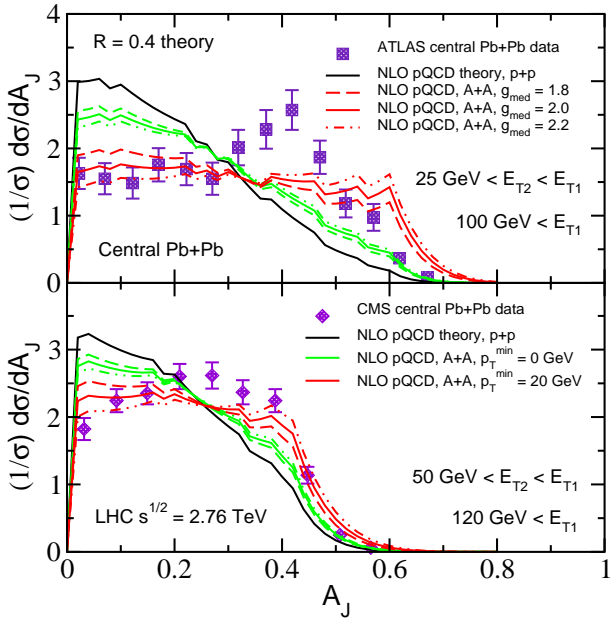}
}
\caption{ Di-jet asymmetry distributions for different coupling strength
in central Pb+Pb collisions at $\sqrt{s}=2.76$~TeV compared to data. The top panel is for leading
jets of $E_{T\,1}>100$~GeV, subleading jets of $E_{T\, 2}>25$~GeV  (ATLAS~\cite{Aad:2010bu}, $R=0.4$).
The bottom panel is for leading jets of $E_{T\, 1}>120$~GeV, subleading jet of $E_{T\, 2}>50$~GeV
(CMS~\cite{Chatrchyan:2011sx}, $R=0.5$).  Black lines are the results for p+p collisions under pQCD theory.
Green lines assume perfect jet/background separation and are denoted $p_{T}^{\min}=0$~GeV.
Red lines, denoted $p_{T}^{\min}=20$~GeV, show the consequences of the ambiguity in the
jet/background separation  or the diffusion of the in-medium parton shower energy
away from the jet axis due to collisional processes.}
\label{AJAA1}
\end{figure}

\begin{figure}
\vskip0.04\linewidth
\centerline{
\includegraphics[width = 0.9\linewidth]{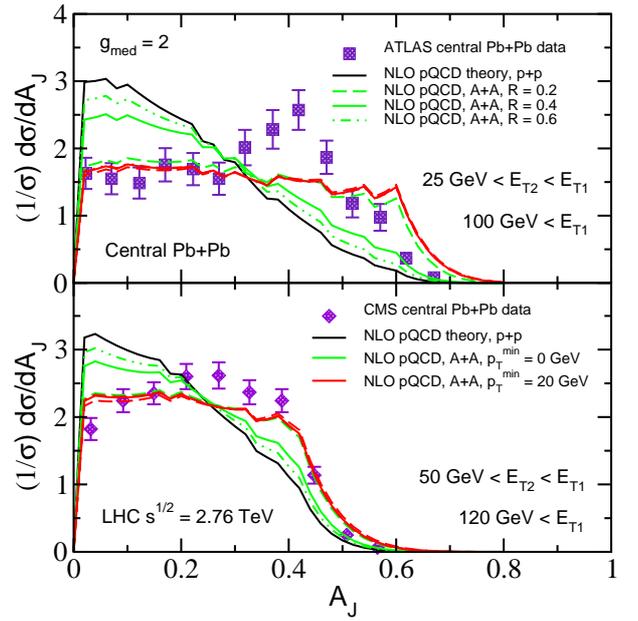}
}
\caption{ Same as Fig~\ref{AJAA1}, but for a fixed jet-medium interaction strength $g_{\rm med}=2$.
Dot-dashed, solid, and dashed lines correspond to $R=0.6, \, 0.4, \,0.2$ respectively.
Note that if some of the subtracted ``background'' is in fact related to the medium-induced
parton shower or energy diffuses away from the jet axis due to collisional processes,
the enhanced asymmetry becomes radius-independent.}
\label{AJAA2}
\end{figure}

Next, the di-jet asymmetry is computed using Eq.~(\ref{ajcalc}) and  compared  to
recent experimental results published by ATLAS~\cite{Aad:2010bu} and
CMS~\cite{Chatrchyan:2011sx} in Fig.~\ref{AJAA1} and Fig.~\ref{AJAA2}. The ATLAS
and  CMS experiments apply different transverse energy cuts for their leading and
subleading jets: ($E_{T\,1} > 100$~GeV,   $ E_{T\,1} >E_{T\,2} > 25$~GeV)  and
($E_{T\,1} > 120$~GeV, $ E_{T\,1} >E_{T\,2} > 50$~GeV), respectively. Two different calculations
are performed and shown in the top and bottom panels of Figs.~\ref{AJAA1} and~\ref{AJAA2}.
In all cases the baseline p+p asymmetry at $\sqrt{s}=2.76$~TeV is shown by a solid black line.
Note that we present the normalized asymmetry result $ (1/\sigma) d\sigma/dA_J $.

Fig.~\ref{AJAA1} shows the sensitivity of the di-jet asymmetry to the coupling strength
$g_{\rm med}$ of the jets to the QGP medium. We use one jet size $R=0.4$ (we will vary $R$ later).
This $p_T^{\min} = 0$~GeV case is represented by green curves. Qualitatively, we find a
broadening of the di-jet asymmetry distribution but the dependence of this broadening to the variation
of $g_{\rm med} =  1.8, \, 2, \, 2.2$ is quite modest. Quantitatively, approximately 1/3 to 1/2
of the extra  broadening  that experiments attribute to jet interactions in matter can
be described.   Next, we consider collisional interactions of the parton shower or an  ambiguity
in the separation of the jets from the background in central Pb+Pb reactions
at the LHC. This is represented and modeled by $p_T^{\min} = 20$~GeV and shown by red lines in
Fig.~\ref{AJAA1}. In this case, a significantly larger broadening is observed. However,
only part of it is due to radiative jet quenching processes. The remainder may be due to removing the soft
particles originating from the medium-induced parton shower  inside the jet
as uncorrelated soft background. In this sense, our result is similar to the argument
presented in~\cite{Cacciari:2011tm} that background fluctuations can generate much of the
asymmetry observed by ATLAS and CMS. Alternatively, there is  the possibility that this
energy is dissipated outside the cone through collisional processes~\cite{source}, something
that we will investigate further in the future. This scenario is also modeled
by $p_T^{\min} = 20$~GeV and in this case we have made quantitative connection to
theoretical simulations~\cite{source}. We finally point out that even if all the energy
associated with the medium-induced parton shower is removed, the resulting $A_J$
distribution is flat. Specifically, a peak in this distribution at finite $A_J = 0.3 - 0.4$
is not compatible with realistic jet quenching calculations\footnote{At the recent
QM2011 conference, after the results presented in this manuscript were posted to the archive,
the ATLAS Collaboration presented a re-analysis of their $A_J$ measurement with improved
background subtraction. The peak at $A_J \simeq 0.4$ is no longer present and the new
distribution is consistent with flat up to the kinematic bound.}.
This is easy to be understood
from the $ {\cal O}( \alpha_s^3 ) $ results presented here. The interested reader can analyze Fig.~\ref{3DRdijet}
and see that the flat $A_J$ distribution is related to the very broad  approximately
constant $R_{AA}^{\rm 2-jet}$.

Fig.~\ref{AJAA2} shows the dependence of the di-jet asymmetry on the jet radius $R$ and the
the momentum cut $p_T^{\min}$ with a fixed coupling strength $g_{\rm med}=2$.
Note that for $p_T^{\min} = 0$~GeV there is a significant dependence on the jet size.
For $p_T^{\min} = 20$~GeV,
the dependence on the radius is practically eliminated. This observation is compatible with the
comment by the ATLAS Collaboration~\cite{Aad:2010bu} that their asymmetry measurement
has little sensitivity to the choice of $R$ over a wide range of cone sizes $R = 0.2-0.6$.
It once again stresses the importance of considering the interaction between the full parton shower
and the QGP beyond the medium-induced parton splitting processes~\cite{source} and/or
the ambiguity in the experimental jet/background separation~\cite{Cacciari:2008gd}.

\section{Summary and Conclusions}
\label{conclude}

Jet production and modification in high-energy nucleus-nucleus collisions have been proposed
as  new ways to unravel the properties of the hot and dense QCD medium and to elucidate
the mechanisms of in-medium parton shower formation~\cite{Vitev:2008rz,jets,Neufeld:2010fj}.
Measurements of jets in heavy ion collisions have now become
available~\cite{Salur:2010qk,Jia:2010hg,Ploskon:2009zd}. With this in mind,
in this Letter we presented first $ {\cal O}(\alpha_s^3)$, $ {\cal O}(\alpha_s^2 \alpha_s^{\rm rad}) $
results for the single and double inclusive jets production
rates in p+p and central Pb+Pb collisions at the LHC at center-of-mass energies per nucleon-nucleon
pair $\sqrt{s_{NN}} = 7$~TeV and 2.76~TeV, respectively. We placed  emphasis on the
good agreement between the perturbative QCD theory and the experimental measurements
in p+p collisions. The importance of reliable $ {\cal O}( \alpha_s^3 ) $ calculation cannot be under
emphasized, since for tagged jets  and di-jets an inaccurate description of the p+p baseline at
LO can produce even qualitatively incorrect jet quenching predictions~\cite{Neufeld:2010fj}.

We first addressed the suppression of inclusive jet production in central Pb+Pb collisions
at the LHC and found that it is dominated by final-state inelastic parton interactions in the
QGP. Cold nuclear matter effects~\cite{Vitev:2008vk,Vitev:2007ve,Neufeld:2010dz}, even though
larger than those for inclusive $Z^0$ production~\cite{Chatrchyan:2011ua}, still do not contribute
significantly to the attenuation of jets for $ R \leq  0.4 $. On the other hand, there is a
clear dependence of $R_{AA}^{\rm 1-jet}$ on the coupling strength between the jet and the medium
$g_{\rm med}$. We found that for small radii the suppression of jets matches onto the suppression
of leading particles~\cite{jets}, here represented by the ALICE charged hadron measurement.
Cold nuclear matter effects can be practically eliminated by taking the ratio of the cross
sections at two different jet sizes $\sigma(R_1)/\sigma(R_2)$, but in this case there is
a sensitivity to the non-perturbative hadronization effects. We used an extrapolation
of the ATLAS parametrization  of non-perturbative effects~\cite{:2010wv}
to show that for small radii $R\sim 0.2$ they can alter the cross section ratios by as
much as 50\%. It is important in the future to extend the experimental measurements to
lower $E_T$ and better constrain the magnitude of the non-perturbative effects.

Theoretically a clean calculation that focuses on the medium modification of jets
is easy to perform. Experimentally, the situation is much more complicated in heavy ion
reactions where there is an enormous soft background that may contribute $\sim 100$~GeV
to the $E_T$ of an $R=0.5$ jet at the LHC. The problem of this background separation is both
technical and conceptual since there is no unified first-principle understanding of
heavy ion dynamics at all momentum scales.
In our calculation we investigated the experimental consequences of this
ambiguity by simulating a subtraction of a radius-dependent $\langle \delta E_T \rangle$
from the medium-induced parton shower as a part of the soft uncorrelated background.
Such amount of energy $\langle \delta E_T \rangle$  may also easily be transferred
from a 100~GeV shower to the medium through collisional processes. The first possibility
{\em emulates} energy loss. The second possibility {\em is} the collisional energy loss of
the parton shower that falls inside the jet cone radius $R$. We used the calculation
in Ref~\cite{source} to constrain its magnitude.
We found that such subtraction wipes out the jet size dependence of the quenching
observables.

We also presented the first calculation of the di-jet modification $R_{AA}^{\rm 2-jet}$
in heavy ion reaction and found that it is characterized by a broad suppression
region near $E_{T\,1} = E_{T\,2}$ and strong enhancement for $E_{T\,1} \gg E_{T\,2}$
or $E_{T\,1} \ll E_{T\,2}$. The resulting $ {\cal O}( \alpha_s^3 ) $ di-jet cross section was used to evaluate
the di-jet asymmetry $A_J$. Assuming perfect jet/background separation we found that
less than 1/2 of the broadening can be accounted for the radiative jet quenching calculation.
We found little sensitivity of the $A_J$ observable to the strength of the coupling
between the jet and the medium and practically no dependence on the non-perturbative
hadronization corrections. On the other hand, we found  a clear dependence on the jet
cone size $R$ that can be correlated to the suppression of inclusive jets.  If part of
the energy of the medium-induced parton shower is additionally lost either through
subtraction as soft background or through collisional processes,
significantly larger enhancement in asymmetric jet production can be obtained for
radii of moderate size $R=0.4-0.5$. For small radii, such as $R=0.2$, practically all
of the asymmetry is due to radiative processes. Finally, the full calculation has
no sensitivity to the choice of the cone radius $R$.

The ambiguity of jet/background
separation and the diffusion of the in-medium parton shower energy through
collisional processes~\cite{source} may play an important role in the generation
of $A_J$. Recovery of the lost energy at large angles outside of the jet
cone~\cite{Chatrchyan:2011sx} would tend to favor the latter scenario.
We are currently looking for new observables~\cite{Neufeld:2012df} that have reduced sensitivity to
QGP background fluctuations but retained clear dependence on the amount of collisional
energy loss of the parton shower. We finally
point out that a statistical peak in the asymmetry distribution at finite
$A_J = 0.4- 0.5$ is likely not related to the physics of jet quenching.

To summarize, significant experimental progress in jet physics in heavy ion collisions
has been made to date. However, taken at face value, the PHENIX and STAR preliminary
results in Au+Au and Cu+Cu
collisions at RHIC~\cite{Salur:2010qk,Ploskon:2009zd} and the ATLAS and CMS jet
results in Pb+Pb collisions at the LHC~\cite{Aad:2010bu,Chatrchyan:2011sx} do not allow, at
present, to construct a consistent picture of jet modification in heavy ion reactions.
Specifically, the pronounced  dependence of the inclusive jet $R_{AA}$  on the
cone radius $R$ at RHIC is inconsistent with argued independence of the di-jet asymmetry
$A_J$ on $R$. Such independence can easily arise from the inherent ambiguity in the
separation of the jet from the enormous soft background in heavy ion reactions. It can also
be related to the diffusion of a part of the in-medium parton shower energy away
from the jet axis due to collisional processes. The ATLAS and CMS di-jet measurements
have played an important role in emphasizing the importance
of jet studies in heavy ion reactions at the LHC. However, a more comprehensive
suite of measurements of inclusive jet and di-jet modification and further developments in
theory are necessary to find a consistent picture of jet production and modification
in dense QCD matter. \\

\noindent {\bf Acknowledgments:}
We thank D. Soper for helpful discussion. This research is  supported by the US
Department of Energy, Office of Science, under
Contract No. DE-AC52-06NA25396 and in part by the LDRD program at LANL, and by
the Ministry of Education of China with the Program NCET-09-0411,
by Natural Science Foundation of China with Project No. 11075062,
NSF of Hubei with Project No. 2010CDA075 and CCNU with Project No. CCNU09A02001.



%% References


\bibliographystyle{model1-num-names}
\bibliography{<your-bib-database>}




\end{document}